\documentclass{article}

\usepackage{arxiv}

\usepackage[utf8]{inputenc} 
\usepackage[T1]{fontenc}    
\usepackage{hyperref}       
\usepackage{url}            
\usepackage{booktabs}       
\usepackage{amsfonts}       
\usepackage{nicefrac}       
\usepackage{microtype}      
\usepackage{lipsum}		
\usepackage{graphicx}
\usepackage{natbib}
\usepackage{doi}

\usepackage{mathtools,upgreek,eucal,mathrsfs,enumitem,amsthm,amsmath}
\usepackage{times,epsfig,makeidx,amsfonts,amsmath,dcolumn,multirow,amssymb,colortbl,bm,url}
\usepackage{algorithm}
\usepackage[]{algorithmic}
 \usepackage[flushleft]{threeparttable}
\usepackage{subfig}

\newtheorem{scenario}{Scenario}

\newcommand{\blambda}{\bm{\lambda}}

\newcommand{\bPsi}{\bm{\Psi}}
\newcommand{\bgamma}{\bm{\gamma}}
\newcommand{\btheta}{\bm{\theta}}
\newcommand{\bbeta}{\bm{\beta}}

\newcommand{\bkappa}{\bm{\kappa}}
\newcommand{\bPhi}{\bm{\Phi}}

\newcommand{\Bmid}{\,\Big\vert \,}

\newcommand{\bb}{{\bf b}}
\newcommand{\bs}{{\bf s}}
\newcommand{\bw}{{\bf w}}
\newcommand{\bx}{{\bf x}}

\newcommand{\bM}{{\bf M}}

\title{A tractable Bayesian joint model for longitudinal and survival data}

\author{ \href{https://orcid.org/0000-0003-3764-0397}{\includegraphics[scale=0.06]{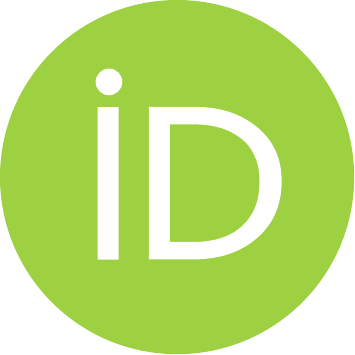}\hspace{1mm}Danilo Alvares}\\
Department of Statistics\\
Pontificia Universidad Cat{\'o}lica de Chile\\
	Macul, Chile \\
	\texttt{dalvares@mat.uc.cl} \\
	\And
	\href{https://orcid.org/0000-0001-7183-8407}{\includegraphics[scale=0.06]{orcid.pdf}\hspace{1mm}Francisco Javier Rubio} \\
	Department of Statistical Science\\
	University College London \\
	London, UK\\
	\texttt{acukfjr@ucl.ac.uk} \\
}

\renewcommand{\headeright}{Technical Report}
\renewcommand{\undertitle}{Technical Report}
\renewcommand{\shorttitle}{A tractable Bayesian joint model}

\hypersetup{
pdftitle={A tractable Bayesian joint model},
pdfsubject={stat.ME, stat.AP},
pdfauthor={Alvares, Rubio},
}

\begin{document}
\maketitle

\begin{abstract}
We introduce a numerically tractable formulation of Bayesian joint models for longitudinal and survival data. The longitudinal process is modelled using generalised linear mixed models, while the survival process is modelled using a parametric general hazard structure. The two processes are linked by sharing fixed and random effects, separating the effects that play a role at the time scale from those that affect the hazard scale. This strategy allows for the inclusion of non-linear and time-dependent effects while avoiding the need for numerical integration, which facilitates the implementation of the proposed joint model. We explore the use of flexible parametric distributions for modelling the baseline hazard function which can capture the basic shapes of interest in practice. We discuss prior elicitation based on the interpretation of the parameters. We present an extensive simulation study, where we analyse the inferential properties of the proposed models, and illustrate the trade-off between flexibility, sample size, and censoring. \textcolor{black}{We also apply our proposal to two real data applications in order to demonstrate the adaptability of our formulation both in univariate time-to-event data and in a competing risks framework.} The methodology is implemented in \texttt{rstan}.
\end{abstract}


\section{Introduction} \label{sec:intro}

In medical statistics, it is common to come across scenarios where patients are followed-up for a period of time (typically, until death or a censoring time point), and some biomarkers, patient characteristics, or treatment information are recorded at different time points over this period. This produces a combination of longitudinal and survival information about each individual. Historically, both processes have been analysed separately. For example, modelling time-to-event data is typically done by using hazard-based regression models. These include the Cox Proportional Hazard (PH) model, \cite{cox:1972} which assumes that the covariates have an effect at the hazard scale; Accelerated Failure Time (AFT) models, \cite{kalbfleisch:2011} which assume that the covariates have a direct effect on the survival time; Accelerated Hazard (AH) models, which assume that the effect of the covariates is only on the time scale of the hazard function; as well as other general hazard (GH) structures that generalise the PH, AFT, and AH assumptions \citep{chen:2001}. See \cite{rubio:2019S} for a general overview of such models. The longitudinal process is typically modelled using Generalised Linear Mixed Models (GLMMs), which allow for modelling repeated and correlated observations (see, e.g. \cite{mcculloch:2008} for a general overview). It has been shown that combining both the longitudinal and survival processes represents a powerful tool for incorporating the information in both processes. Joint modelling of longitudinal and survival processes has been extensively discussed in recent literature. We refer the reader to literature \cite{rizopoulos:2012,elashoff:2016,furgal:2019,alsefri2020} for reviews on this sort of models. A common strategy in joint models consists of linking the survival and the longitudinal processes by means of including shared parameters on the models for the covariates. This allows for incorporating a number of statistical modelling tools already available in the literature, such as using flexible parametric models using splines for modelling the hazard or the cumulative hazard functions (see, \cite{brilleman:2019} for a recent review on these methods), while the longitudinal process can be modelled using any techniques developed for GLMMs. Applications of joint models abound in a number of areas of medical statistics \citep{rue:2017,hickey:2018,mauff:2020}.

In this paper, we propose a numerically tractable and interpretable alternative formulation of joint models, where we allow the longitudinal process to be modelled using GLMMs, while the survival process is specified through a parametric general hazard structure. This formulation allows for a direct interpretation of the parameters, as they are formulated at the hazard scale, as well as a separation of the roles of the parameters that affect the time scale, from those that affect the hazard scale. Another appealing aspect of the proposed formulation of joint models is numerical tractability, as the evaluation of the hazard and cumulative hazard functions does not require numerical integration, allowing for a tractable implementation of the likelihood and posterior distribution functions. We discuss several choices for the baseline hazard that are able to capture a variety of shapes of the hazard function. We discuss prior elicitation, where the general idea is to use weakly information priors for shape parameters while, for regression parameters, we consider g-priors \citep{zellner1986} in order to ameliorate potential overfit of those variables modelled using splines. Thus, the proposed joint models can be coupled with a number of general-purpose MCMC samplers. We provide an implementation of these models in \texttt{rstan} \citep{stan:2020} and show a good performance of this sampler in our simulation study and applications. We provide an extensive simulation study that illustrates the performance of our joint specification as well as the trade--off between using flexible assumptions for modelling the baseline hazard and non-linear effects, with sample size and censoring. In addition, we use a data set on AIDS patients \citep{goldman1996} to illustrate our methodology in a standard joint model context. We also present another real data example, using the SANAD study \citep{marson:2007}, where the survival process contains competing risks, emphasising the flexibility of our formulation to be coupled with a variety of scenarios. The rest of the paper is organised as follows. In Section \ref{sec:jm}, we present the formulation of the joint model and discuss the interpretation of the parameters. In Section \ref{sec:bi}, we present the likelihood function in a general framework, and discuss prior elicitation for the case where the longitudinal process is modelled using a linear mixed model (LMM), which is the model used later in the real data applications. In Section \ref{sec:sim}, we discuss an extensive simulation study and indicate how to simulate from the proposed joint model. In Sections \ref{sec:appl1} and \ref{sec:appl2}, we illustrate the proposed methodology with two real data applications in the contexts of univariate time-to-event and competing risks, respectively. Finally, in Section \ref{sec:disc}, we present a brief discussion of the proposal in this paper and conclude with some practical advice and potential directions for further research. Additional results, including summaries from the simulation study, alternative models in the application, as well as technical details are presented in the Supplementary Material. R code is available at: \url{www.github.com/daniloalvares}.

\section{The Joint Model} \label{sec:jm}

\subsection{Longitudinal model: generalised linear mixed model}

The longitudinal component of the proposed joint model is specified through a GLMM \citep{mcculloch:2008}. Let $y_{ij} = y_i(t_j)$ be the response variables associated to the $i$th individual, $i=1,\dots,n$, measured at time $t_j$, $j=1,\dots,n_i$. Let $\bx_i \in {\mathbb R}^p$ be a vector of individual covariates corresponding to the $i$th individual. Define the conditional distribution of $y_{ij}$ given $\bPsi_{1i}$ (parameters and random effects) as a member of the Exponential family:
\begin{eqnarray}\label{eq:long}
y_{ij} \mid \bPsi_{1i} &\stackrel{ind.}{\sim} &  f_L(y_{ij}\mid \bPsi_{1i}), \nonumber\\
f_L(y_{ij} \mid \bPsi_{1i}) &=& \exp \left\{ \dfrac{y_{ij}\xi_{ij} - \varphi(\xi_{ij})}{\tau^2} - c(y_{ij},\tau) \right\}, \nonumber\\
E[y_{ij} \mid \bPsi_{1i}] &=& \mu_{ij}, \nonumber\\
g(\mu_{ij}) &=& \tilde{\beta}_0 + {\bs}_i^{\top}\bbeta + \tilde{\bx}_i^{\top}\bgamma P_{1}(t_{ij}) + b_{0i} + (\tilde{\beta}_1 + b_{1i}) P_{2}(t_{ij}),
\end{eqnarray}
where $\varphi$ is a known function, the conditional mean of $y_{ij}$ given $\bPsi_{1i}$ is related to $\xi_{ij}$ via the identity $\mu_{ij} = \dfrac{\partial \varphi(\xi_{ij})}{\partial \xi_{ij}}$, the conditional variance of $y_{ij}$ given $\bPsi_{1i}$ is $\tau^2\dfrac{\partial^2 \varphi(\xi_{ij})}{\partial \xi_{ij}^2}$, and $g$ is the link function. Regarding the model on the mean $\mu_{ij}$, $\tilde{\beta}_{0}$ is the intercept, $\tilde{\beta}_{1}$ is the time-dependent slope; ${\bs}_i = (s_{i1}^{\top},\dots, s_{ip}^{\top})^{\top} \in {\mathbb R}^{p q}$, where $q=\sum_{i=1}^p q_i$, $q_m$, $m=1,\dots,p$, is the dimension of $s_{im}$, and $s_{im}$ is a spline expansion of $x_{im}$ (for continuous variables, or simply $x_{im}$ for categorical variables or variables with a linear effect); $\bbeta \in {\mathbb R}^{pq}$ are the corresponding regression coefficients; $\tilde{\bx}_i \subseteq \bx_i \in{\mathbb R}^{\tilde{p}}$, $\tilde{p}\leq p$, is a vector of individual time-dependent covariates, and $\bgamma\in{\mathbb R}^{\tilde{p}}$ are the corresponding regression coefficients; $b_{0i}$ and $b_{1i}$ are the random effects, which represent a random intercept and a random slope. This formulation thus allows for the inclusion of linear and non-linear effects by using a spline expansion of the corresponding covariates. $P_{1}(t_{ij})$ and $P_{2}(t_{ij})$ represent polynomial expansions, which indicate the functional dependence of time of the time-dependent covariates $\tilde{\bx}_i $, the slope $\tilde{\beta}_1$, and the random slopes $b_{1i}$. These can be, for instance, a $B-$spline basis polynomial expansion, or simply the identity function \citep{rizopoulos:2012}. In practice, it is often assumed a linear relationship, unless the individual trajectories are suspected to be non-linear. We assume that the random effects, $\bb_i = (b_{0i},b_{{1}i})^{\top}$, given $\Sigma$, follow a joint bivariate normal distribution with zero mean and variance-covariance matrix $\Sigma$. This family of mixed models include linear mixed models, Poisson mixed models, Negative Binomial mixed models, binary mixed models, among others.

\subsection{Survival model: general hazard structure}

In this section, we discuss the model for the survival process, in which we adopt a general hazard (GH) structure \citep{chen:2001,rubio:2019S}. Let $h_0(\cdot \mid \btheta)$ be a parametric baseline hazard function, with parameter $\btheta \in \Theta \subset {\mathbb R}^d$. Define the hazard function:
\begin{equation}\label{eq:hazGH}
h(t \mid \bPsi_{2i}) = h_{0}\left(t \exp\left\{ {\bw}_i^{\top}{\bkappa} +  \alpha_1\left( \tilde{\bx}_i^{\top}\bgamma + b_{1i} \right) \right\} \Bmid \btheta\right) \exp\left\{\tilde{\bw}_i^{\top}\tilde{\bkappa} + \bs_i^{\top}\blambda+ \alpha_0 b_{0i} \right\},
\end{equation}
where $t>0$ represents the time; $\bPsi_{2i}^{\top} = (\btheta^{\top}, \bgamma^\top, \bkappa^\top, \tilde{\bkappa}^\top, \blambda^{\top},  \bb_{i}^{\top},\alpha_0,\alpha_1)$ denotes the full vector of model parameters; ${\bw}_i $ and $\tilde{\bw}_i$ are $r-$ and $\tilde{r}-$dimensional vectors of additional covariates, affecting the time-scale and the hazard-scale, respectively, which may only be available for the survival process (\emph{i.e.}~this formulation allows for the inclusion of different variables in the longitudinal and survival models); $\bkappa$ and $\tilde{\bkappa}$ are the regression coefficients associated to ${\bw}_i $ and  $\tilde{\bw}_i$, respectively; $\blambda$ are regression coefficients for the a expansion $\bs_i$; $\alpha_0\in{\mathbb R}$ and $\alpha_1\in{\mathbb R}$ are the parameters linking the longitudinal and the survival processes, often called \textit{association parameters}. 

The hazard structure \eqref{eq:hazGH} separates the roles of the time-dependent effects (which appear in the argument of the baseline hazard and, consequently, affect directly the time scale) from those effects on the hazard scale, which appear multiplying the baseline hazard \citep{rubio:2019S}. This is, the hazard structure \eqref{eq:hazGH} can be used to account for time-dependent effects as well as effects that either increase or decrease the hazard level while the link with the longitudinal process also explicitly separates these roles \citep{rubio:2019S}. We see this as an advantage of this formulation as it helps to identify the need for connecting the two processes via time-dependent and/or proportional hazard effects. This hazard model can be directly used in more complex scenarios such as competing risks models, which we illustrate in our real data application.
Another appealing feature of this hazard structure is that the corresponding cumulative hazard can be written in closed-form, thus avoiding the need for numerical integration, as:
\begin{eqnarray}\label{eq:chazGH}
H(t \mid \bPsi_{2i}) &=& H_{0}\left(t \exp\left\{ {\bw}_i^{\top}{\bkappa} +  \alpha_1\left(   \tilde{\bx}_i^{\top}\bgamma  +  b_{1i}  \right) \right\} \Bmid \btheta\right)\exp\left\{\tilde{\bw}_i^{\top}\tilde{\bkappa} + \bs_i^{\top}\blambda+ \alpha_0 b_{0i}  - \left[   {\bw}_i^{\top}{\bkappa} +  \alpha_1\left(  \tilde{\bx}_i^{\top}\bgamma  +  b_{1i}  \right) \right] \right\},
\end{eqnarray}
where $H_0(\cdot \mid \btheta)$ is the cumulative baseline hazard of $h_0(\cdot \mid \btheta)$. This allows for a tractable implementation of the likelihood and posterior distributions, which in turns allows this joint model formulation to be coupled with any general-purpose MCMC sampler. 

This model specification is identifiable provided that the baseline hazard is not the hazard associated to a Weibull distribution \citep{chen:2001,rubio:2019S}, since in this case the AFT, PH, and AH models coincide and it becomes impossible to distinguish the effects in the argument of the baseline hazard from those multiplying the baseline hazard. We do not consider this to be a big limitation as similar hazard shapes can be obtained with other distributions, as discussed next, and model selection tools can be used to identify simpler hazard structures.

A natural extension of the hazard structure \eqref{eq:hazGH} consists of using more than one link parameter $\alpha_1$, for instance, in cases where the vector $\tilde{\bx}_i$ contains variables in very different scales. However, we argue that, in practice, this is not often required, and estimating more than one scaling parameter would require larger samples. This logic is also in line with the classical formulation of joint models \citep{rizopoulos:2012}, where only one link parameter is used. We also note that, under the formulation in \eqref{eq:hazGH}, we are connecting the survival and longitudinal processes through the random effects and the time-dependent effects. It is, of course, possible to link them through the time-invariant effects ${\bs}_i^{\top}\bbeta$ as follows, 
\begin{equation*}
h(t \mid \bPsi_{2i}) = h_{0}\left(t \exp\left\{ {\bw}_i^{\top}{\bkappa} +  \alpha_1\left(  \tilde{\bx}_i^{\top}\bgamma  +  b_{1i}  \right) \right\} \Bmid \btheta\right) \exp\left\{\tilde{\bw}_i^{\top}\tilde{\bkappa} + \alpha_0 (b_{0i} + {\bs}_i^{\top}\bbeta)\right\},
\end{equation*}
which reduces the number of parameters by omitting $\blambda$. However, one limitation of this approach is that it assumes that the scaling factor is the same for all covariates, which may be in different scales or may contain a combination of categorical and continuous variables. Nonetheless, it remains as an alternative formulation for modelling the survival process.

\subsubsection*{Baseline hazard function}

The choice of the parametric baseline hazard function is crucial as this determines the hazard shapes the survival model \eqref{eq:hazGH} can capture. For instance, the log-normal hazard function is unimodal (up-then-down), while the Gamma hazard function can be increasing or decreasing. There exist other (three-parameter) distributions that can capture the basic shapes of the hazard (increasing, decreasing, unimodal, and bathtub), such as the Exponentiated Weibull, Generalised Gamma, and Power Generalised Weibull distributions. However, it is important to consider that an efficient estimation of the parameters of these distributions typically requires larger sample sizes, and that high censoring rates or early administrative censoring (short follow-up) may also be detrimental in estimating shape parameters (specially those that control the tail behaviour) of flexible parametric distributions \citep{rubio:2019S,rossell:2019}. We consider four baseline hazard candidates: Log-normal, Gamma, Power Generalised Weibull (PGW), and Generalised Gamma (GG), based on their numerical tractability and flexibility. The PGW and GG distributions contain three parameters (a scale parameter, and two shape parameters). These distributions offer similar levels of tractability and flexibility  \citep{jones:2015}. Expressions for the PGW and GG  probability density functions (pdf), survival functions, and hazard functions are presented in Sections A1-A2 in the Supplementary Material.

\section{Bayesian inference} \label{sec:bi}

\subsection{Likelihood function}

The likelihood function of the full parameter vector and random effects of the joint model \eqref{eq:long}--\eqref{eq:hazGH} is given by:
\begin{equation}\label{eq:like}
f( \text{Data} \mid \bPsi ) =  \prod_{i=1}^{n} \prod_{j=1}^{n_i} f_L(y_{ij} \mid \bPsi_{1i})  \prod_{i=1}^{n}f_S(t_{i} \mid \bPsi_{2i})  ,
\end{equation}
where $\bPsi = (\bPsi_{1}^\top,\bPsi_{2}^\top)^\top =(\bbeta^{\top}, \tilde{\bbeta}^{\top}, \bgamma^\top, \tau, \bb_{1}^{\top},\ldots, \bb_{n}^{\top}, \btheta^{\top}, \bkappa^{\top}, \tilde{\bkappa}^\top, \blambda^{\top}, \alpha_0, \alpha_1)^{\top}$ denotes the full parameter vector and random effects; $f_L(y_{ij} \mid \bPsi_{1i})$ denotes the conditional pdf of $y_{ij}$ given $\bPsi_{1i}$ described in \eqref{eq:long}; and $f_S(t_{i} \mid \bPsi_{2i})$ is the contribution of the $i$th time-to-event to the likelihood function. For example, for linear mixed models, which we describe in the next section, $f_L(y_{ij} \mid \bPsi_{1i})$ can be the normal density with mean $\mu_{ij}$ and variance $\sigma^2$. The contribution of the survival time $t_{i}$ is described by:
$$
f_S(t_{i} \mid \bPsi_{2}) =
     \begin{cases}
       h(t_{i} \mid \bPsi_{2i}) S(t_{i} \mid \bPsi_{2i}), & \text{if exact lifetime,} \\
       S(t_{i} \mid \bPsi_{2i}),  & \text{if right-censored observation,} \\
       1 - S(t_{i} \mid \bPsi_{2i}),  & \text{if left-censored observation,} \\
       S(t_{i,L} \mid \bPsi_{2i}) - S(t_{i,R} \mid \bPsi_{2i}), & \text{if interval-censored observation,} 
     \end{cases}
$$
where $h(t_{i} \mid \bPsi_{2i})$ is the hazard function \eqref{eq:hazGH} and $S(t_{i} \mid \bPsi_{2i}) = \exp\left\{-H(t_{i} \mid \bPsi_{2i})\right\}$ represents the survival function derived from the cumulative hazard introduced in \eqref{eq:chazGH}.

Although we will focus on Bayesian inference for the parameters, we point out that the marginal likelihood function of the parameters $\bPsi^{\ast} = \bPsi_{-\left\{\bb_1,\ldots,\bb_n\right\}}$ can be written as follows:
\begin{eqnarray*}\label{eq:marglike}
f(\text{Data} \mid \bPsi^{\ast},  \Sigma) &=& \prod_{i=1}^{n} \int \left[\prod_{j=1}^{n_i} f_L(y_{ij} \mid \bPsi_{1i})\right] f_S(t_{i} \mid \bPsi_{2i}) \phi(\bb_{i} \mid \Sigma) \ d\bb_{i},
\end{eqnarray*}
where $\phi(\bb_{i} \mid \Sigma)$ is the bivariate normal density with zero mean and variance-covariance matrix $\Sigma$. Thus, the evaluation of the marginal likelihood function, under our joint model formulation, only requires numerical integration with respect to the distribution of the random effects.

\subsection{Prior distributions} \label{ssec:priors}

In this section, we provide general guidelines about prior choice for the parameters of the proposed joint models. We identify weakly informative priors for the parameters of this new class of joint models, but we acknowledge the possibility of using alternative priors. 

We define the prior distributions for the more particular case of linear mixed models (LMMs), which are the main interest in our applications. However, these can be extended to GLMMs by adapting the priors on the variance of the response variable, accordingly \citep{li:2018}. Consider the LMM:
\begin{equation*}\label{eq:LMM}
y_{i}(t) = \tilde{\beta}_0 + {\bs}_i^{\top}\bbeta + \tilde{\bx}_i^{\top}\bgamma P_{1}(t_{ij}) + b_{0i} + (\tilde{\beta}_1 + b_{1i}) P_{2}(t_{ij}) + \epsilon_{i}(t).
\end{equation*}

The residual errors are assumed conditionally independent and identically distributed as $(\epsilon_{i}(t) \mid \sigma^{2}) \sim N(0,\sigma^{2})$. In order to avoid concerns about the propriety of the posterior distribution \citep{rubio:2018}, we adopt a proper prior specification. For the parameters $(\tilde{\beta}_0, \tilde{\beta}_1, \bgamma, \Sigma )$, we adopt weakly informative priors:
\begin{eqnarray*}
\tilde{\beta}_j &\sim& N(0,\phi_{\tilde{\beta_j}}^2),\,\,\,\,\, j=0,1,\\
\bgamma &\sim& N_{\tilde{p}}(0,\bPhi_{\tilde{\bgamma}}), \\
\sigma^2 &\sim& \text{Inv-Gamma}(0.01,0.01),\\
\sigma_j^2 &\sim& \text{Inv-Gamma}(0.01,0.01),\,\,\,\,\, j=1,2,\\
\dfrac{\rho + 1}{2} &\sim& \text{Beta}(a_0,b_0),
\end{eqnarray*}
where $\rho \in (-1,1)$ is the correlation between $b_{0i}$ and $b_{1i}$.  The variance hyperparameters, $\phi_{\tilde{\beta_j}}^2$ and the diagonal of $\bPhi_{\tilde{\bgamma}}$, are assumed to be large in order to reflect vague prior information. Alternatively, one could choose a half-Cauchy prior for the variance parameters \citep{rubio:2018}. For the parameters $\bbeta$ and $\blambda$, which represent regression coefficients associated to covariates that may contain spline expansions, we consider the following prior specification that penalises overfit:
\begin{eqnarray*}
 \pi(\bbeta \mid \sigma^2) &=& \prod_{I_O} N({\bbeta}_j;0,\phi_{\bbeta_j}^2) \prod_{I_S} N({\bbeta}_j; 0,g_{\bbeta} \bM_j \sigma^2), \\
 \pi(\blambda \mid \eta^2) &=&  \prod_{I_O} N({\blambda}_j;0,\phi_{\blambda_j}^2) \prod_{I_S} N({\blambda}_j; 0,g_{\blambda} \bM_j \eta^2),
\end{eqnarray*}
where $I_O = \{j: \bx_{ij} = \bs_{ij}, \text{for all } i\}$ and $I_S = \{j: \bx_{ij}  \neq \bs_{ij}, \text{for all } i\}$ indicate the indexes of the variables expressed in the original scale and in a spline basis expansion, respectively, $\bM_j = (\tilde{S}_j^{\top}\tilde{S}_j)^{-1}$, and $\tilde{S}_j$ are the design matrices associated to spline basis expansions of the covariates $\bx_{ij}$. This is, if the vector $\bs_i$ contains spline expansions of $\bx_i$, we adopt g-priors \citep{zellner1986}; otherwise, we adopt weakly informative priors where the variance hyperparameters $\phi_{\bbeta_j}^2$ and $\phi_{\blambda_j}^2$ are assumed to be large. The hyperparameters $g_{\bbeta} = n/q$ and $g_{\blambda} = n/q$ \citep{rossell:2019}, assuming that all spline basis expansions have the same degree $q$ (which can be easily relaxed, if necessary), induce a mild penalty that shrinks the parameters towards zero and help prevent overfitting (see  \cite{rossell:2019} for a discussion on this point and other choices of these hyperparameters). We keep a relatively simple prior choice in this paper, but we acknowledge the possibility of including other shrinkage priors that carry heavier penalties on model complexity \citep{simpson:2017,rossell:2019}. We emphasise that those priors can also be included in our approach and numerical implementation.

For the parameter $\btheta$ in baseline hazards, we consider the following priors:
\begin{enumerate}
\item {\it Log-normal} (LN). For the scale parameter, say $\eta>0$, we adopt a weakly informative prior $\eta \sim \text{half-Cauchy}(0,s_{\eta})$, in the sense that it is a heavy tailed prior \citep{rubio:2018}. For the log-location parameter, we assume $\mu \sim N(0,\phi_{\mu}^2)$, where the hyperparameter $\phi_{\mu}^2$ is assumed to be large.
\item {\it Gamma}. For the scale and shape parameters, we adopt a weakly informative prior $\eta \sim \text{half-Cauchy}(0,s_{\eta})$ and $\nu \sim \text{half-Cauchy}(0,s_{\nu})$, respectively.
\item {\it Power Generalised Weibull} (PGW). For the scale, shape, and power parameters, $(\eta,\nu,\delta)$, we adopt weakly informative priors specified as $\eta \sim \text{half-Cauchy}(0,s_{\eta})$, $\nu \sim \text{half-Cauchy}(0,s_{\nu})$, and $\delta \sim \text{Gamma}(1.83,0.65)$. The prior on the parameter $\delta$ represents an approximation to the weakly informative prior $BTV(1,1)$ obtained with the method proposed in \cite{dette:2018}. A full description of this prior is presented in Section A2 in the Supplementary Material.
\item {\it Generalised Gamma} (GG). Similar to PGW specification, $\eta \sim \text{half-Cauchy}(0,s_{\eta})$, $\nu \sim \text{half-Cauchy}(0,s_{\nu})$, and $\delta \sim \text{Gamma}(1.83,0.65)$.
\end{enumerate}
For the association parameters, we adopt the weakly informative priors $\alpha_k \sim N(0,\phi_{{\alpha_k}}^2)$, $k=0,1$, where the variance hyperparameters are assumed to be large. We point out that shrinkage priors  \citep{andrinopoulou:2016} could also be considered for these parameters, which might be useful to enforce parsimony in the link between the longitudinal and survival processes.

\section{Simulations} \label{sec:sim}

\subsection{Simulating from the joint model}

We now describe a method to simulate from the proposed joint model \eqref{eq:long}--\eqref{eq:hazGH}. This method basically requires simulating the random effects first, followed by the simulation of the survival process using the GH structure \eqref{eq:hazGH}, and finally simulating the longitudinal process \eqref{eq:long}. The steps for the simulation procedure are described in Algorithm \ref{alg:simJM}. We can see that simulating from the survival model GH is relatively simple, in contrast to other joint models discussed in the literature, provided that one can simulate times-to-event from the baseline model. Thus, the choice of tractable baseline hazards model also facilitates the simulation of the joint model. Regarding the distribution of the distance between repeated observations (DDBRO), we have several scenarios of practical importance. For instance, in medical scenarios with periodic consultations, the DDBRO would be equidistant. In more complex scenarios, this distance might be random, for instance visits to the hospital due to some treatment or illness complication, which are also recorded and monitored; or even a combination of periodic and random visits to the hospital. Our formulation allows for the inclusion of all of these types of DDBRO. Censored survival times can be induced in the standard way, by either inducing administrative censoring or simulating random censoring points. 
\begin{algorithm}[ht]
\caption{Simulation from the proposed joint model \label{alg:simJM}}
For each individual $i=1,\dots,n$, and for given values of the parameters and the design matrix:
\begin{algorithmic}
\STATE \textbf{Random Effects}. Simulate $\bb_i\sim N_{2}(0,\Sigma)$.
\STATE  \textbf{Survival Process}.
In order to simplify notation, let us denote:
\begin{eqnarray*}
A &=& \exp\left\{ {\bw}_i^{\top}{\bkappa} + \alpha_1\left( \tilde{\bx}_i^{\top}\bgamma + b_{1i}  \right) \right\},\\
B &=& \exp\left\{ \tilde{\bw}_i^{\top}\tilde{\bkappa} + \bs_i^{\top}\blambda+ \alpha_0 b_{0i}  - \left[   {\bw}_i^{\top}{\bkappa} +  \alpha_1\left(  \tilde{\bx}_i^{\top}\bgamma + b_{1i}  \right) \right]  \right\}.
\end{eqnarray*}
The individual survival function is $S(t \mid \bPsi_{2i}) = \exp\left[-H(t \mid \bPsi_{2i})\right]$, we can apply the probability integral transform directly to obtain:
\begin{equation*}
t_{i} = \dfrac{F_0^{-1}\left[ 1 -\exp\left\{ \dfrac{\log(1-u_{i})}{B} \right\} \Bmid \btheta \right]}{A},
\end{equation*}
where $F_0$ is the cumulative distribution function associated to the baseline hazard $h_0$, and $u_{i} \sim U(0,1)$.
\STATE \textbf{Longitudinal Process}. Once a simulated time-to-event $t_i$ is obtained from the previous step, specify the distribution of the distance between the repeated observations (e.g.~equidistant or random). This produces the time points $t_{ij}$, $j=1,\dots,n_i$, at which the repeated observations are recorded. The longitudinal process simulation is thus obtained by plugging-in the corresponding values of the parameters and covariates in $\mu_{ij}$, and simulating from the corresponding GLMM based on \eqref{eq:long}.
\end{algorithmic}
\end{algorithm}

\subsection{Simulation study}

In this section, we conduct an extensive simulation study where we present the performance of the proposed joint model and estimation methods. More specifically, we illustrate the parameter estimation, ability to recover the baseline hazard shapes, as well as the effect of sample size and censoring rates on inference. For the survival process, we consider a scenario where the available variables are age at diagnosis, sex, and the presence of comorbidities. This setting is common in population studies in cancer epidemiology \citep{rubio:2019B}. We analyse the following simulation scenarios, in increasing order of complexity.

\begin{scenario}
The longitudinal model:
\begin{equation*}
y_i(t) = \tilde{\beta}_0 + \tilde{\beta}_1 t +  \beta_1 \,\text{sex}_i + \beta_2 \, \text{age}_i + b_{0i} + b_{1i}t + \epsilon_i(t).
\end{equation*}

The survival process:
\begin{equation*}
h(t \mid \bPsi_{2i}) = h_{0}\left(t \exp\left\{ \alpha_{1} b_{1i} \right\} \Bmid \btheta\right) \exp\left\{ \tilde{\kappa}_1 \text{comorb}_i  +  \lambda_1 \,\text{sex}_i + \lambda_2 \, \text{age}_i +\alpha_0 b_{0i} \right\}.
\end{equation*}
\end{scenario}

In addition, we consider \textbf{Scenario 0} where we simulate the model described in \textbf{Scenario 1} but we fit the following joint model. The longitudinal model:
\begin{equation*}
y_i(t) = \tilde{\beta}_0 + \tilde{\beta}_1 t + \beta_1 \, \text{sex}_i + \beta_2 \, \text{age}_i + b_{0i} + b_{1i}t + \epsilon_i(t).
\end{equation*}

The survival process:
\begin{equation*}
h(t \mid \bPsi_{2i}) = h_{0}\left(t \mid \btheta\right) \exp\left\{ \tilde{\kappa}_1 \text{ comorb}_i  +  \lambda_1 \,\text{sex}_i + \lambda_2 \, \text{age}_i +\alpha_0 b_{0i} \right\}.
\end{equation*}

\begin{scenario}
The longitudinal model:
\begin{equation*}
y_i(t) = \tilde{\beta}_0 + \tilde{\beta}_1 t + \gamma_1 \left\{t  \, \text{age}_i \right\} + \beta_1 \,\text{sex}_i + \beta_2 \, \text{age}_i + b_{0i} + b_{1i}t + \epsilon_i(t).
\end{equation*}

The survival process:
\begin{equation*}
h(t \mid \bPsi_{2i}) = h_{0}\left(t \exp\left\{  \alpha_{1}\left(\gamma_1 \, \text{age}_i +b_{1i} \right) \right\} \Bmid \btheta\right)  \exp\left\{ \tilde{\kappa}_1 \text{\ comorb}_i +  \lambda_1 \,\text{sex}_i + \lambda_2 \, \text{age}_i +\alpha_0 b_{0i} \right\}.
\end{equation*}
\end{scenario}

\begin{scenario}
The longitudinal model:
\begin{equation*}
y_i(t) = \tilde{\beta}_0 + \tilde{\beta}_1 t + \gamma_1 \left\{t  \, \text{age}_i \right\} +  \beta_1 \, \text{sex}_i + g_1(\text{age}_i) +  b_{0i} + b_{1i}t + \epsilon_i(t).
\end{equation*}

The survival process:
\begin{equation*}
h(t \mid \bPsi_{2i}) = h_{0}\left(t \exp\left\{ \alpha_{1}\left(\gamma_1 \, \text{age}_i +b_{1i} \right) \right\} \Bmid \btheta\right)  \exp\left\{ \tilde{\kappa}_1 \text{comorb}_i  + \lambda_1 \,\text{sex}_i + g_2(\text{age}_i)  +\alpha_0 b_{0i} \right\},
\end{equation*}
where $g_1$ and $g_2$ are B-spline expansions of the variable $\text{age}_i$ of degree $q=3$.
\end{scenario}
Thus, in Scenarios 2-3, the variable ``age'' represents a time-dependent effect both on the longitudinal and the survival process. This can be seen as it appears multiplying $t$ in the longitudinal process, while it affects the time-scale directly on the survival process. In Scenario 3, the variable ``age'' also includes non-linear effects in both processes. 
The variable ``age'' is simulated from a mixture of uniform variables with probabilities $0.25$ for the age group $(30,65)$, $0.35$ for the age group $(65,75)$, and $0.4$ for the age group $(75,85)$. This variable is then centered at $70$ and scaled by a factor of $10$. The variables ``comorbidity'' and ``sex'' are simulated from a Binomial distribution with probability parameter $0.5$. The parameter values in each of these scenarios are presented in Section A3 in the Supplementary Material. These values are selected, together with the administrative censoring points, in order to obtain $5\%$ and $35\%$ censoring rates, which will allow us to assess the effect of censoring. We also consider two sample sizes in each scenario, $n=200$ and $n=500$, in order to evaluate the effect of sample size. The residual errors are assumed conditionally independent and identically distributed as $(\epsilon_{i}(t) \mid \sigma^{2}) \sim N(0,\sigma^{2})$. 

For each simulation scenario, we simulate $N=100$ data sets and obtain $2,000$ posterior samples of the parameters of the corresponding joint models using \texttt{rstan}. We apply a burn-in period of $1,000$ iterations as well as a thinning period of $5$ iterations to these posterior samples, for a total of $200$ posterior samples. Under this configuration, we have observed convergence of the posterior samples. The number of Monte Carlo iterations is based on a trade-off between CPU time while trying to minimise the Monte Carlo error. Simulations were performed on an iMac with 3.3 GHz Quad-Core Intel Core i7, 16 GB RAM, macOS Catalina. 

The results from this simulation study are presented in Sections A4 and A5 in the Supplementary Material. Tables A7--A38 display summaries of the posterior samples. These tables present the averages of the posterior means, posterior medians, $2.5\%$ posterior quantiles, and $97.5\%$ posterior quantiles for each of the posterior samples in each scenario. We observe a good performance of the Bayesian point estimators, as these means are close to the true values of the parameters. There is a clear effect of the sample size and censoring rates on the accuracy of the estimates. Unsurprisingly, the larger the sample, the more accurate the estimates. A similar conclusion is obtained for the censoring rates, as we can see that lower censoring rates produce more accurate estimates. Model complexity is an important factor (for instance, log-normal \textit{vs.}~PGW baseline, or linear model \textit{vs.}~splines model), as we can see that the more complex the model is, the wider the credible intervals are, which also interacts with the effects of sample size and censoring rates. A more interesting conclusion is observed for the length of follow-up, as reducing the length of follow-up has a marked effect on the accuracy of the estimates of the parameters of the survival model. In particular, the estimation of the shape parameters in the PGW distribution suffers from early administrative censoring as these parameters control the tail of the distribution, while early administrative censoring removes information about the tails of the distribution. This phenomenon remains even after increasing the sample size, indicating that a longer follow-up might be as important as increasing the sample size if the aim is to improve the accuracy of the estimates. In addition, Figures A3--A22 in the Supplementary Material show the baseline hazards associated to the plugging posterior median estimators as well as the posterior predictive baseline hazards (which are defined as the ratio of the posterior predictive probability density function and the posterior predictive survival function). We observe a similar situation about the effect of the censoring rate, sample size, and model complexity on the ability to recover the shape of the baseline hazard. Another interesting result, obtained from comparing Scenarios 0 and 1, is that not including $\alpha_1 b_{1i}$ (i.e.~not sharing the random slopes) induces a bias in the estimation of $\alpha_0$. However, in order to be able to estimate the parameter $\alpha_1$ accurately, a longer follow-up or a larger sample size is necessary. Finally, comparing Scenarios 1 and 2, we can see that it is easier to estimate the link parameters when fixed and random time-dependent effects are combined (for 3-parameter baseline hazard), in contrast to the case when only random time-dependent effects are considered. This applies to all sample sizes and censoring rates. In this case, the use of an appropriate simpler model (in the sense that it can capture the true hazard shape) improves the estimation of the link parameters. 

We conduct additional simulation studies (Section A6) where we assess the effect of higher censoring rates ($60\%$ censoring) and for binary longitudinal outcomes (which is modelled using a logistic mixed effects model). Results are shown in Tables A39--A58 in the Appendix. The conclusions are the same as those in the previous paragraph in terms of the interplay between sample size, censoring, and the flexibility of the baseline hazard and the functional form of the regression model. This section also illustrates the fact that results are the same for negative correlations and negative regression coefficients (as the role of the parameters remains the same). Finally, Section A6 also presents a simulation study using a Generalised Gamma baseline hazard instead of PGW. The results are comparable to those obtained in the PGW case, however, the need for using special functions for evaluating the GG hazard and cumulative hazard functions has a large cost in terms of computing times as the posterior sampling is slowed down by a factor of 7, compared to the PGW model, despite the efficient implementation of these functions discussed in the appendix. Thus, although equally flexible, the GG model also carries a larger computational cost, which is more apparent in the Bayesian framework where the evaluation of the posterior needs to be done thousands of times to obtain a posterior MCMC sample. 

Overall, this simulation study provides some guidelines (and a warning) about the use of flexible models with many parameters when there are high censoring rates or small samples. Thus, in practice, it is recommended to compare simple models against more complex models using formal model selection tools. We will illustrate this idea in the following section.

\section{The AIDS study: univariate time-to-event joint model} \label{sec:appl1}

The \texttt{aids} data consists of 467 patients with advanced human immunodeficiency virus infection during antiretroviral treatment who had failed or were intolerant to zidovudine therapy \citep{goldman1996}. This data set is publicly available in the R package \texttt{JM} \citep{JM} and the main objective is to identify associations between the time to death and the following covariates: \texttt{CD4}: CD4 cells count (longitudinal biomarker); \texttt{prevOI}: a factor with levels \texttt{AIDS} denoting previous opportunistic infection (AIDS diagnosis) at study entry, and \texttt{noAIDS} denoting no previous infection; \texttt{drug}: a factor with levels \texttt{ddC} denoting zalcitabine and \texttt{ddI} denoting didanosine; \texttt{gender}: a factor with levels \texttt{female} and \texttt{male}; \texttt{AZT}: a factor with levels \texttt{intolerance} and \texttt{failure} denoting AZT intolerance and AZT failure, respectively.

For simplicity, we used only the baseline variable \texttt{prevOI}, coded as 0 (\texttt{noAIDS}) and 1 (\texttt{AIDS}), to illustrate our methodological approach. In addition, the longitudinal \texttt{CD4} variable was transformed by applying the square root. This selection is also based on a preliminary study on the importance of these variables on the longitudinal outcome of interest.

\subsection{Model 1 ($M_1$)} \label{ssec:model11}

Our first proposal specifies the longitudinal model with fixed effects for the intercept ($\tilde{\beta}_0$), slope ($\tilde{\beta}_1$), and \texttt{prevOI} ($\beta$); random effects for the intercept ($b_{0}$) and slope ($b_{1}$); and a time-dependent effect for the \texttt{prevOI} covariate ($\gamma$). Mathematically, we can represent $y_i(t)$ as the $\sqrt{\text{\texttt{CD4}}}$ at time $t$ for patient $i=1,\ldots,n=467$ and therefore the longitudinal model is defined by:
\begin{eqnarray}
y_i(t) &=& \tilde{\beta}_0 + \tilde{\beta}_1 t + \gamma \, \text{prevOI}_i \, t + \beta \,\text{prevOI}_i + b_{0i} + b_{1i} \, t + \epsilon_i(t), \label{app:long11}
\end{eqnarray}
where $\epsilon_i(t) \stackrel{ind.}{\sim} N(0,\sigma^{2})$ is the random error at time $t$ and we assume that the random effects, $\bb_i = (b_{0i},b_{1i})^{\top}$, given $\Sigma$, follow a joint bivariate normal distribution with zero mean and variance-covariance matrix $\Sigma$.

The hazard function at time $t$ is defined as:
\begin{eqnarray} 
h(t \mid \bPsi_{2i}) &=& h_{0}\left(t \exp\left\{  \alpha_{1}\left(\gamma \, \text{prevOI}_i +b_{1i} \right) \right\} \Bmid \btheta\right) \exp\left\{ \lambda \,\text{prevOI}_i +\alpha_{0} b_{0i} \right\}, \label{app:surv1}
\end{eqnarray}
where $h_{0}$ is a baseline hazard function that will be specified here as Log-normal, Gamma, Generalised Weibull, and Generalised Gamma (see Section \ref{ssec:priors} for more details of these specifications); $\alpha_{0}$ and $\alpha_{1}$ denote the association parameters; and $\lambda$ is the regression coefficient for \texttt{prevOI}.

\subsection{Model 2 ($M_2$)} \label{ssec:model12}

The longitudinal model of our second proposal is slightly different from the specification of \eqref{app:long11}, as now we do not include the time-dependent effect for the \texttt{prevOI} covariate. Then, the linear mixed model (LMM) is expressed by:
\begin{eqnarray}
y_i(t) &=& \tilde{\beta}_0 + \tilde{\beta}_1 t + \beta \,\text{prevOI}_i + b_{0i} + b_{1i} \, t + \epsilon_i(t). \label{app:long12}
\end{eqnarray}

In this joint approach specification, the survival model shares only the random effects:
\begin{eqnarray}
h(t \mid \bPsi_{2i}) &=& h_{0}\left(t \exp\left\{ \alpha_{1}b_{1i} \right\} \Bmid \btheta\right) \exp\left\{ \lambda \,\text{prevOI}_i +\alpha_{0} b_{0i} \right\}. \label{app:surv2}
\end{eqnarray}

\subsection{Model 3 ($M_3$)} \label{ssec:model13}

Our third proposal models the longitudinal process as in \eqref{app:long12}, but shares only the random intercept:
\begin{eqnarray}
h(t \mid \bPsi_{2i}) &=& h_{0}\big(t \mid \btheta \big) \exp\left\{ \lambda \,\text{prevOI}_i +\alpha_{0} b_{0i} \right\}. \label{app:surv3}
\end{eqnarray}

The prior distributions for the parameters and hyperparameters of models $M_{1}$, $M_{2}$ and $M_{3}$ are specified as in Section~\ref{ssec:priors}.

\subsection{Bayesian model comparison} \label{ssec:comparison1}

Suppose we have $m$ Bayesian models, say $M_{1},\ldots,M_{m}$, to be compared. So, the relative plausibility of a particular model $M_{v}$ given its prior probability and the evidence from the data is quantified by the so-called {\it posterior model probability} (PMP) \citep{berger2005}, defined as follows:
\begin{eqnarray}
\text{PMP}_{v} = P(M_{v} \mid \text{Data}) = \frac{P(\text{Data} \mid M_{v})P(M_{v})}{\sum_{j=1}^{m}P(\text{Data} \mid M_{j})P(M_{j})}, \quad v=1,\ldots,m, \label{app:pmp}
\end{eqnarray}

\noindent where again we assumed that the models are equally probable \textit{a priori}.

\subsection{Results} \label{ssec:results1}

We start the analysis by comparing the joint models $M_{1}$, $M_{2}$ and $M_{3}$ using the following baseline hazards: Log-normal, Gamma, Power Generalised Weibull, and Generalised Gamma. Table~\ref{tab:comparison1} shows the approximate calculation of posterior model probabilities, obtained with the \texttt{post\_prob} function, available in the R package \texttt{bridgesampling} \citep{gronau2020}.

\begin{center}
\begin{table}[htb]
\centering
\caption{Baseline hazard model comparison based on posterior model probability (PMP). \label{tab:comparison1}}
\begin{tabular*}{500pt}{@{\extracolsep\fill}ccccc@{\extracolsep\fill}}
\toprule
Model   & Log-normal & Gamma & Power Generalised Weibull & Generalised Gamma \\
\midrule
 1 & $0$    & $0$ & $0$ & $0$ \\
\midrule
 2 & $0$ & $\bm{0.9944}$ & $0$ & $0.0001$ \\
\midrule
 3 & $0$ & $0.0011$ & $0.0044$ & $0$ \\
\bottomrule
\end{tabular*}
\end{table}
\end{center}

The results indicate $M_{2}$ with the Gamma baseline hazard as the best model. Table~\ref{tab:model1} shows a posterior summary for this model. The last column of this table contains the posterior probability that the corresponding parameter is positive. A probability equal to $0.5$ indicates that a positive value of the parameter is equally likely than a negative one.

\begin{center}
\begin{table}[htb] 
  \centering
  \caption{Posterior summary for model $M_{2}$ with the Gamma baseline hazard specification. \label{tab:model1}}
  \begin{tabular*}{500pt}{@{\extracolsep\fill}ccccccc@{\extracolsep\fill}}
    \toprule
    Interpretation          & Parameter            & Mean    & Median   & $2.5\%$  & $97.5\%$ & $P(\cdot > 0 \mid \text{Data})$ \\
    \midrule
    intercept               & $\tilde{\beta}_{0}$  & $3.111$   &  $3.111$   &  $2.986$   &  $3.242$  & $1.000$ \\
    slope                   & $\tilde{\beta}_{1}$  & $-0.042$  &  $-0.042$  &  $-0.050$  &  $-0.034$ & $0.000$ \\
    prevOI (AIDS)           & $\beta$              & $-0.910$  &  $-0.909$  &  $-1.067$  &  $-0.754$ & $0.000$ \\
    intercept RE variance   & $\sigma^{2}_{b_{0}}$ & $0.582$   &  $0.582$   &  $0.500$   &  $0.678$  & -- \\
    slope RE variance       & $\sigma^{2}_{b_{1}}$ & $0.002$   &  $0.002$   &  $0.001$   &  $0.002$  & -- \\
    RE correlation          & $\rho$               & $0.040$   &  $0.039$   &  $-0.143$  &  $0.231$  & $0.655$ \\
    error variance          & $\sigma^{2}$         & $0.134$   &  $0.134$   &  $0.120$   &  $0.150$  & -- \\		
		\midrule
    prevOI (AIDS)              & $\lambda$    & $1.622$   &  $1.613$   &  $1.181$   &  $2.108$  & $1.000$ \\
    intercept RE association   & $\alpha_{0}$ & $-0.935$  &  $-0.935$  &  $-1.207$  &  $-0.666$ & $0.000$ \\
    time association           & $\alpha_{1}$ & $-25.681$ &  $-25.587$ &  $-44.110$ &  $-8.053$ & $0.002$ \\
    Gamma scale                & $\eta$       & $33.081$  &  $32.007$  &  $20.172$  &  $52.348$ & -- \\
    Gamma shape                & $\nu$        & $1.760$   &  $1.740$   &  $1.440$   &  $2.195$  & -- \\
    \bottomrule		
    \end{tabular*}
    \begin{tablenotes}
      \item prevOI: Previous Opportunistic Infection at study entry. \; RE: Random Effect.
    \end{tablenotes}	
\end{table}
\end{center}

The first seven parameters in Table~\ref{tab:model1} refer to the longitudinal model for the $\sqrt{\text{CD4}}$. In particular, it is important to note that the posterior mean of the amount of CD4 (in square root scale) to patients with previous opportunistic infection (AIDS diagnosis) at study entry was less than that of patients with no previous infection by E$(\beta \mid \text{Data})=-0.91$ units.

As expected, the group of patients with previous opportunistic infection at study entry has a higher risk of death, E$(\lambda \mid \text{Data})=1.615$. In addition, the association parameters ($\alpha_{0}$ and $\alpha_{1}$) indicate that by having a low CD4 amount at study entry (intercept) or even decreasing this amount throughout the study (slope), the risk of death increases.

\begin{figure}[htb!]
    \centering
    \subfloat[]{{\includegraphics[width=7.5cm]{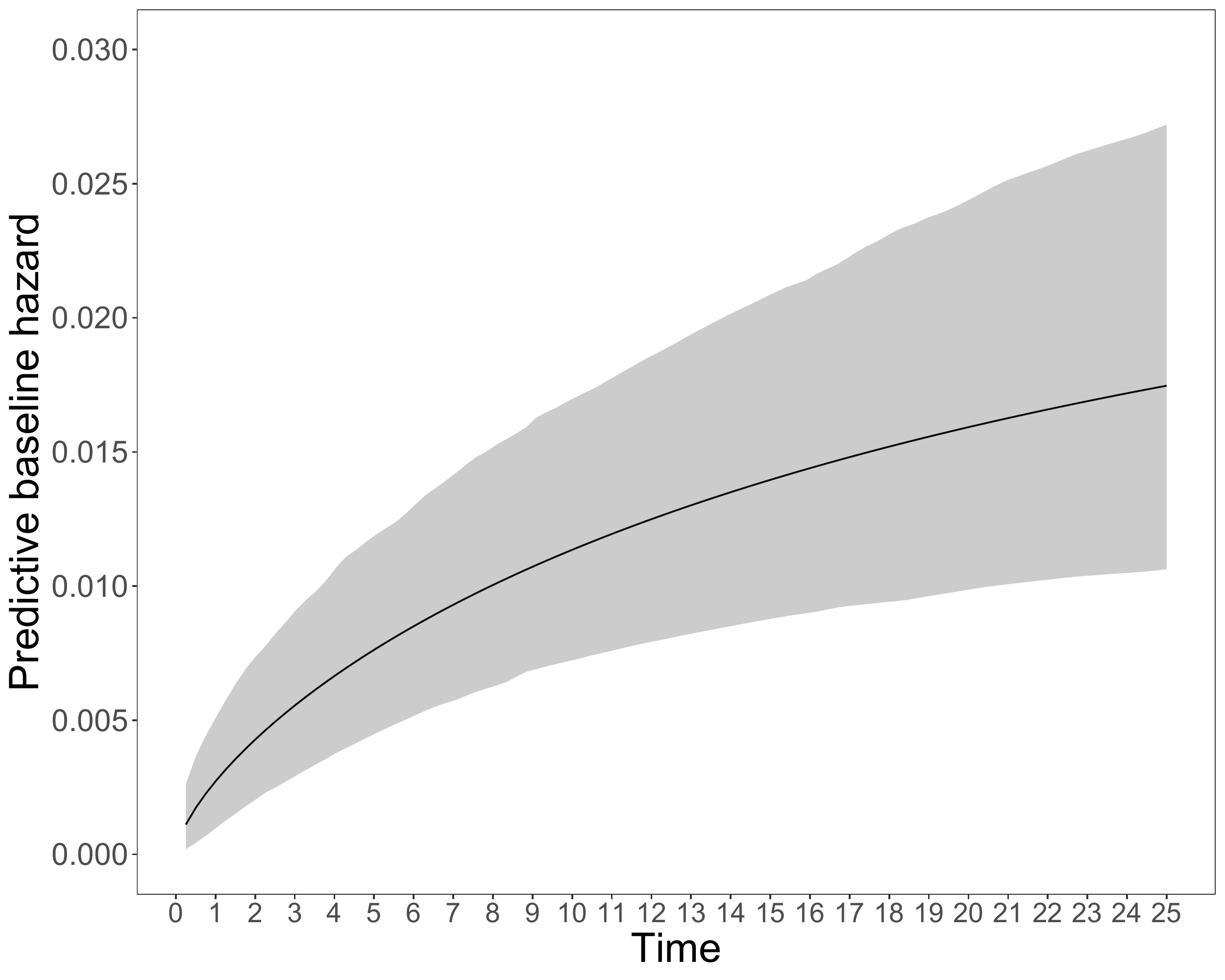}}}	
    \quad \quad \quad \quad
    \subfloat[]{{\includegraphics[width=7.5cm]{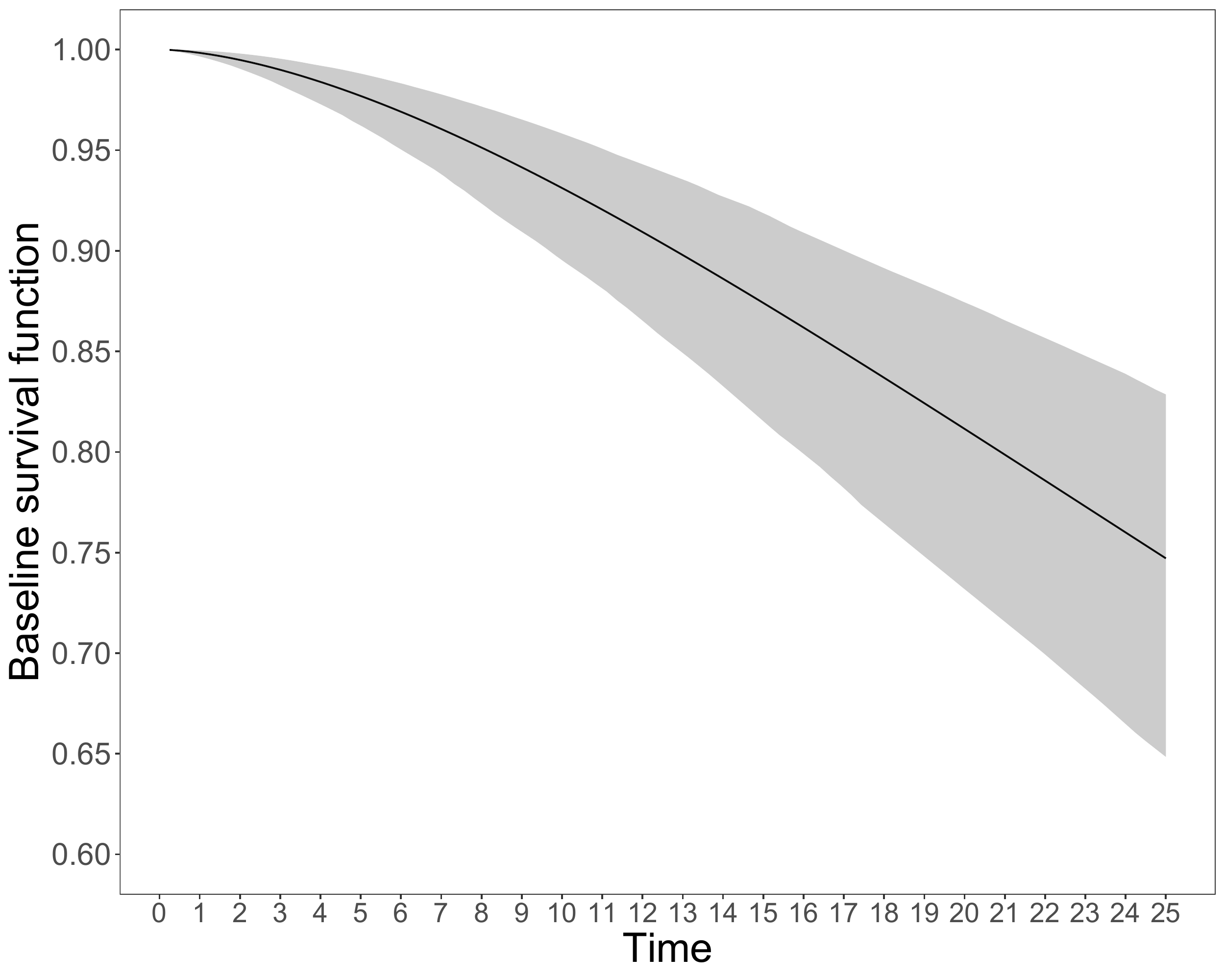}}}
    \caption{Gamma predictive (a) baseline hazard and (b) survival functions, and their respective $95\%$ credible intervals for model $M_{2}$.}
    \label{fig:Gammamodel2}
\end{figure}

Considering the following parameterisation of the Gamma probability distribution function $f_{0}(t)=\zeta^{\nu}t^{\nu-1}e^{-\zeta t}/\Gamma(\nu)$, Figure~\ref{fig:Gammamodel2} shows the plots of the predictive baseline hazard, $h_{0}(t)=f_{0}(t)/S_{0}(t)$, and survival, $S_{0}(t)$, and their respective 95\% credible interval using the posterior samples of $\zeta$ and $\nu$ from model $M_{2}$.

\section{The SANAD study: competing risks joint model} \label{sec:appl2}

The SANAD (Standard and New Anti-epileptic Drugs) study, designed and analysed by \cite{marson:2007}, is an unblinded randomised controlled trial in hospital-based outpatient clinics conducted between 1998 and 2006 in the UK. Partial data from this study is publicly available in the R package \texttt{joineR} \citep{joineR}, where the main objective is to investigate the time to treatment failure (here defined as the withdrawal of a randomised drug or addition of another) based on a standard anti-epileptic drug (carbamazepine, CBZ) and a new drug (lamotrigine, LTG). The time to treatment failure can occur due to two competing events: inadequate seizure control (ISC) or unacceptable adverse effects (UAE). Table~\ref{tab:summary} shows a brief summary of the baseline covariates and time to events for each competing event.

\begin{center}
\begin{table}[htb]%
\centering
\caption{Competing event status, baseline covariates and time to events. \label{tab:summary}}
\begin{tabular*}{500pt}{@{\extracolsep\fill}lccc@{\extracolsep\fill}}
\toprule
 & Censored & ISC & UAE \\
\midrule
    $n$                                   & $391$         & $120$         & $94$ \\
    \texttt{gender}: Female $|$ Male         & $173 \; | \; 218$ & $52 \; | \; 68$   & $37 \; | \; 57$ \\
    \texttt{treat}: CBZ $|$ LTG              & $179 \; | \; 212$ & $55 \; | \; 65$   & $58 \; | \; 36$ \\
    \texttt{age}: Mean (SD) \, [in years]    & $38.4 \; (19.1)$ & $33.6 \; (16.7)$ & $38.9 \; (19.0)$ \\
		\texttt{time}: Median (SD) \, [in years] & $2.2 \; (1.7)$   & $1.3 \; (1.3)$   & $0.5 \; (0.9)$ \\
\bottomrule
\end{tabular*}
\end{table}
\end{center}

Additionally, at each clinical visit, the drug dose of each patient is adjusted if necessary. So, the dose at each visit is a longitudinal marker potentially associated with the time until the events of interest. This time-dependent endogenous covariate, typically modelled through a linear mixed-effects specification, is linked to the competing risks model by means of a joint modelling \citep{elashoff2007}. \cite{williamson2007b,williamson2007a} were the first to analyse this dataset using a competing risk model without longitudinal information. Later, \cite{williamson2008} proposed a joint modelling approach and more recently \cite{hickey:2018} compared different specifications of competing risks joint models for these data.

To model this problem, we propose three flexible specifications for joint models for longitudinal and competing risks data. All proposals model the longitudinal dose variable as a linear mixed model (LMM) and the competing risks data as a cause-specific hazards model \citep{putter2007} using the log-normal baseline specification. The details of each model are described below.

\subsection{Model 1 ($M_1$)} \label{ssec:model21}

Our first proposal specifies the longitudinal model with fixed effects for the intercept ($\tilde{\beta}_0$), slope ($\tilde{\beta}_1$), \texttt{gender} ($\beta_1$), \texttt{treat} ($\beta_2$), and \texttt{age} ($\beta_3$); random effects for the intercept ($b_{0}$) and slope ($b_{1}$); and a time-dependent effect for the \texttt{age} covariate ($\gamma$). Mathematically, we can represent $y_i(t)$ as the drug dose at time $t$ for patient $i=1,\ldots,n=605$ and therefore the longitudinal model is defined by:
\begin{eqnarray}
y_i(t) &=& \tilde{\beta}_0 + \tilde{\beta}_1 t + \gamma \, \text{age}_i \, t + \beta_1 \,\text{gender}_i + \beta_2 \,\text{treat}_i + \beta_3 \, \text{age}_i + b_{0i} + b_{1i} \, t + \epsilon_i(t), \label{app:long21}
\end{eqnarray}
where $\epsilon_i(t) \stackrel{ind.}{\sim} N(0,\sigma^{2})$ is the random error at time $t$ and we assume that the random effects, $\bb_i = (b_{0i},b_{1i})^{\top}$, given $\Sigma$, follow a joint bivariate normal distribution with zero mean and variance-covariance matrix $\Sigma$.

The cause-specific hazard function of the $k$th treatment failure at time $t$ is defined as:
\begin{eqnarray}
h_{k}(t \mid \bPsi_{2i}) &=& h_{k0}\big(t \exp\left\{\alpha_{k1}(\gamma \, \text{age}_i + b_{1i})\right\}  \mid \mu_{k}, \, \eta_{k} \big)\exp\left\{ \lambda_{k1} \,\text{gender}_i + \lambda_{k2} \,\text{treat}_i + \lambda_{k3} \, \text{age}_i +\alpha_{k0} b_{0i} \right\}, \label{app:comprisks1}
\end{eqnarray}
where $h_{k0}$ is a log-normal baseline hazard function with log-location $\mu_{k}$ and scale $\eta_{k}$ parameters; $\alpha_{k0}$ and $\alpha_{k1}$ denote the association parameters; $\lambda_{k1}$, $\lambda_{k2}$ and $\lambda_{k3}$ are the regression coefficients for \texttt{gender}, \texttt{treat} and \texttt{age}; and $k=\text{I},\text{U}$ represent ISC and UAE events, respectively.

\subsection{Model 2 ($M_2$)} \label{ssec:model22}

The longitudinal model of our second proposal is slightly different from the specification of \eqref{app:long21}, as now we do not include the time-dependent effect for the \texttt{age} covariate. Then, the LMM is expressed by:
\begin{eqnarray}
y_i(t) &=& \tilde{\beta}_0 + \tilde{\beta}_1 t + \beta_1 \,\text{gender}_i + \beta_2 \,\text{treat}_i + \beta_3 \, \text{age}_i + b_{0i} + b_{1i} \, t + \epsilon_i(t). \label{app:long22}
\end{eqnarray}

In this joint approach specification, the competing risks model shares only the random effects:
\begin{eqnarray}
h_{k}(t \mid \bPsi_{2i}) &=& h_{k0}\big(t \exp\left\{\alpha_{k1}b_{1i}\right\}  \mid \mu_{k}, \, \eta_{k} \big) \exp\left\{ \lambda_{k1} \,\text{gender}_i + \lambda_{k2} \,\text{treat}_i + \lambda_{k3} \, \text{age}_i +\alpha_{k0} b_{0i} \right\}. \label{app:comprisks2}
\end{eqnarray}

\subsection{Model 3 ($M_3$)} \label{ssec:model23}

Our third proposal models the longitudinal process as in \eqref{app:long22}, but shares only the random intercept:
\begin{eqnarray}
h_{k}(t \mid \bPsi_{2i}) &=& h_{k0}\big(t \mid \mu_{k}, \, \eta_{k} \big) \exp\left\{ \lambda_{k1} \,\text{gender}_i + \lambda_{k2} \,\text{treat}_i + \lambda_{k3} \, \text{age}_i +\alpha_{k0} b_{0i} \right\}. \label{app:comprisks3}
\end{eqnarray}

The prior distributions for the parameters and hyperparameters of models $M_{1}$, $M_{2}$ and $M_{3}$ are specified as in Section~\ref{ssec:priors}.

For these analyses, we coded \texttt{gender} as 0 (Female) and 1 (Male), \texttt{treat} as 0 (CBZ) and 1 (LTG), \texttt{age} was standardised, and the longitudinal \texttt{dose} variable for both groups of drugs were rescaled to have the same range of values. From now on we will refer to the dose as the calibrated dose due to this scale transformation.

\subsection{Bayesian model comparison} \label{ssec:comparison2}

In addition to the posterior model probability (see Section~\ref{ssec:comparison1}), we also used the Bayes factor. Let $M_{v}$ and $M_{j}$ be two Bayesian models competing with each other, then the Bayes factor in favour of $M_{v}$ against $M_{j}$ is defined by:
\begin{eqnarray}
\text{BF}_{vj} = \frac{P(\text{Data} \mid M_{v})}{P(\text{Data} \mid M_{j})} = \frac{P(M_{v} \mid \text{Data})}{P(M_{j} \mid \text{Data})}{\frac{P(M_{v})}{P(M_{j})}}, \label{app:bf}
\end{eqnarray}
where we assumed that $M_{v}$ and $M_{j}$ are equally probable \textit{a priori}, so that $P(M_{v})=P(M_{j})$ and therefore $\text{BF}_{vj}=P(M_{v} \mid \text{Data})/P(M_{j} \mid \text{Data})$. In order to show the comparative results on a more friendly scale, we used log$_{10}$-Bayes factor (LBF) with the interpretations proposed by \cite{kass1995}.

\subsection{Results} \label{ssec:results2}

We started the analysis by comparing the joint models $M_{1}$, $M_{2}$ and $M_{3}$ introduced in previous sections. Table~\ref{tab:comparison2} shows the approximate calculation of Bayes factors and posterior model probabilities, obtained with the \texttt{bf} and \texttt{post\_prob} functions, respectively, available in the R package \texttt{bridgesampling} \citep{gronau2020}.

\begin{center}
\begin{table}[htb]%
\centering
\caption{Model comparison based on posterior model probability (PMP) and log$_{10}$-Bayes factor (LBF). \label{tab:comparison2}}
\begin{tabular*}{500pt}{@{\extracolsep\fill}cccccc@{\extracolsep\fill}}
\toprule
\multicolumn{3}{c}{Posterior model probability} & \multicolumn{3}{c}{Log$_{10}$-Bayes factor} \\
\midrule
  PMP$_{1}$  &  PMP$_{2}$ & PMP$_{3}$ & LBF$_{12}$ & LBF$_{13}$ & LBF$_{23}$ \\
\midrule
  $0$  &  $\bm{1$} &    $0$ &  $-7.20$ &       $19.56$ &    $26.76$  \\
\bottomrule
\end{tabular*}
\end{table}
\end{center}

The results are decisively favorable to model $M_{2}$ and indicates the model $M_{3}$ as the worst option. Table~\ref{tab:model2} shows a posterior summary for model $M_{2}$ and the results for models $M_{1}$ and $M_{3}$ are presented in Section A5 in the Supplementary Material.

\begin{center}
\begin{table}[htb]
  \centering
  \caption{Posterior summary for model $M_{2}$ with a log-normal baseline hazard specification. \label{tab:model2}}
  \begin{tabular*}{500pt}{@{\extracolsep\fill}ccccccc@{\extracolsep\fill}}
    \toprule
    Interpretation          & Parameter            & Mean    & Median   & $2.5\%$  & $97.5\%$ & $P(\cdot > 0 \mid \text{Data})$ \\
    \midrule
    intercept               & $\tilde{\beta}_{0}$  & $1.812$   &  $1.812$   &  $1.674$   &  $1.950$  & $1.000$ \\
    slope                   & $\tilde{\beta}_{1}$  & $0.347$   &  $0.346$   &  $0.299$   &  $0.395$  & $1.000$ \\
    gender (Male)           & $\beta_1$            & $0.089$   &  $0.089$   &  $-0.062$  &  $0.240$  & $0.877$ \\
    treat  (LTG)            & $\beta_2$            & $-0.030$  &  $-0.029$  &  $-0.180$  &  $0.120$  & $0.350$ \\
    age                     & $\beta_3$            & $0.005$   &  $0.005$   &  $-0.068$  &  $0.080$  & $0.553$ \\
    intercept RE variance   & $\sigma^{2}_{b_{0}}$ & $0.745$   &  $0.743$   &  $0.649$   &  $0.851$  & -- \\
    slope RE variance       & $\sigma^{2}_{b_{1}}$ & $0.162$   &  $0.161$   &  $0.127$   &  $0.204$  & -- \\
    RE correlation          & $\rho$               & $0.043$   &  $0.042$   &  $-0.075$  &  $0.163$  & $0.754$ \\
    error variance          & $\sigma^{2}$         & $0.199$   &  $0.198$   &  $0.186$   &  $0.212$  & -- \\		
    \midrule
    & ISC & \multicolumn{5}{c}{} \\
		\midrule
    gender (Male)              & $\lambda_{\text{I}1}$ & $0.039$   &  $0.038$   &  $-0.332$  &  $0.415$  & $0.579$ \\
    treat  (LTG)               & $\lambda_{\text{I}2}$ & $-0.248$  &  $-0.248$  &  $-0.627$  &  $0.132$  & $0.100$ \\
    age                        & $\lambda_{\text{I}3}$ & $-0.254$  &  $-0.253$  &  $-0.464$  &  $-0.046$ & $0.009$ \\
    intercept RE association   & $\alpha_{\text{I}0}$  & $0.181$   &  $0.182$   &  $-0.053$  &  $0.409$  & $0.937$ \\
    time association           & $\alpha_{\text{I}1}$  & $-7.139$  &  $-7.065$  &  $-9.795$  &  $-4.896$ & $0.000$ \\
    log-normal log-location    & $\mu_{\text{I}}$      & $2.959$   &  $2.927$   &  $2.084$   &  $4.014$  & $1.000$ \\
    log-normal scale           & $\eta_{\text{I}}$     & $2.716$   &  $2.708$   &  $2.301$   &  $3.167$  & -- \\
    \midrule		
    & UAE & \multicolumn{5}{c}{} \\
    \midrule
    gender (Male)              & $\lambda_{\text{U}1}$ & $0.196$   &  $0.195$   &  $-0.308$  &  $0.708$  & $0.775$ \\
    treat  (LTG)               & $\lambda_{\text{U}2}$ & $-1.009$  &  $-1.005$  &  $-1.522$  &  $-0.508$ & $0.000$ \\
    age                        & $\lambda_{\text{U}3}$ & $0.139$   &  $0.138$   &  $-0.102$  &  $0.384$  & $0.869$ \\
    intercept RE association   & $\alpha_{\text{U}0}$  & $-1.278$  &  $-1.275$  &  $-1.617$  &  $-0.957$ & $0.000$ \\
    time association           & $\alpha_{\text{U}1}$  & $-12.060$ &  $-11.427$ &  $-21.827$ &  $-5.635$ & $0.000$ \\
    log-normal log-location    & $\mu_{\text{U}}$      & $3.339$   &  $3.295$   &  $2.003$   &  $4.935$  & $1.000$ \\
    log-normal scale           & $\eta_{\text{U}}$     & $2.887$   &  $2.872$   &  $2.383$   &  $3.463$  & -- \\	
    \bottomrule		
    \end{tabular*}
    \begin{tablenotes}
      \item LTG: Lamotrigine. \; RE: Random Effect. \; ISC: Inadequate Seizure Control. \; UAE: Unacceptable Adverse Effects.
    \end{tablenotes}		
\end{table}
\end{center}

The first nine parameters in Table~\ref{tab:model2} refer to the longitudinal model for the calibrated drug dose. E$(\tilde{\beta}_{0} \mid \text{Data})=1.812$ represents the posterior mean of the average value of dose at baseline with its respective underlying posterior variance (intercept random effect) of $0.745$ among patients. The posterior mean dose increment each year was $0.347$ and its posterior mean inter-individual variation (slope random effect) was $0.162$. The mean posterior correlation between the random effects was positive but small ($0.043$). The posterior mean of the amount of dose delivered to men was marginally higher than that of women by E$(\beta_{1} \mid \text{Data})=0.089$ units. The posterior mean of the amount of dose administered to the LTG-treated patient group was insignificantly less than in those CBZ-treated (E$(\beta_{2} \mid \text{Data})=-0.030$). The age of patients was irrelevant in terms of the amount of dose delivered to them. The posterior mean of the error variance was $0.199$.

Regarding the risk of ISC, the patient's gender had virtually no influence, whereas treatment LTG and age produced a decrease in risk of ISC. As the posterior mean of the association parameter for the random intercept was positive, also confirmed by $P(\alpha_{\text{I}0} > 0 \mid \text{Data})=0.937$, and so a high baseline dose leads to a higher risk of ISC. On the other hand, the posterior mean of the association parameter for the random slope was negative, it implies that an increase in dose is associated with a decrease in risk of ISC.

Regarding the risk of UAE, male and/or older patients had a higher risk of UAE, whereas LTG-treated patient group had a reduced risk of UAE compared to those CBZ-treated. For this competing event, a high baseline dose and its increase over time reduced the risk of the patient experiencing UAE, since the posterior mean of the association parameters for the random intercept and slope were distinctly negative. 

In order to visually compare the competing risks, we have used the \textit{predictive baseline hazards} and the \textit{cumulative incidence functions} based on the posterior sample of the log-normal log-location ($\mu$) and scale ($\eta$) parameters for ISC and UAE treatment failures. The predictive baseline hazard function for the $k$th risk is described as:
\begin{equation}
h_{k0}(t \mid \text{Data}) = \frac{f_{k0}(t \mid \text{Data})}{S_{0}(t \mid \text{Data})}, \quad k=\text{I},\text{U}, \label{eq:predhaz}
\end{equation}
where 
\begin{eqnarray*}
f_{k0}(t \mid \text{Data}) &=&  \int_{{\mathbb R}_+} \int_{{\mathbb R}}h_{k0}(t \mid \mu_{k}, \eta_{k}) S_{0}(t \mid \mu_{k}, \eta_{k}) \pi( \mu_{k}, \eta_{k} \mid \text{Data}) \, d \mu_{k} d \eta_{k}, \\
S_{0}(t \mid \text{Data}) &=& \prod_{k=\text{I},\text{U}} \int_{{\mathbb R}_+} \int_{{\mathbb R}} \exp\left\{- \int_0^t h_{k0}(u \mid \mu_{k}, \eta_{k}) \, du \right\} \pi( \mu_{k}, \eta_{k} \mid \text{Data}) \, d \mu_{k} d \eta_{k},
\end{eqnarray*}
are the baseline posterior predictive sub-density function for the $k$th risk and the overall posterior predictive survival function, respectively. The (baseline) posterior predictive cumulative incidence function represents the probability of failure from cause $k$ before time $t$ in the presence of all other possible causes \citep{bakoyannis2012}, and is defined as:
\begin{equation}
F_{k0}(t \mid \text{Data}) = \int_{0}^{t} f_{k0}(u \mid \text{Data}) \, du, \quad k=\text{I},\text{U}. \label{eq:cif}
\end{equation}

\begin{figure}[htb!]
    \centering
    \subfloat[]{{\includegraphics[width=7.5cm]{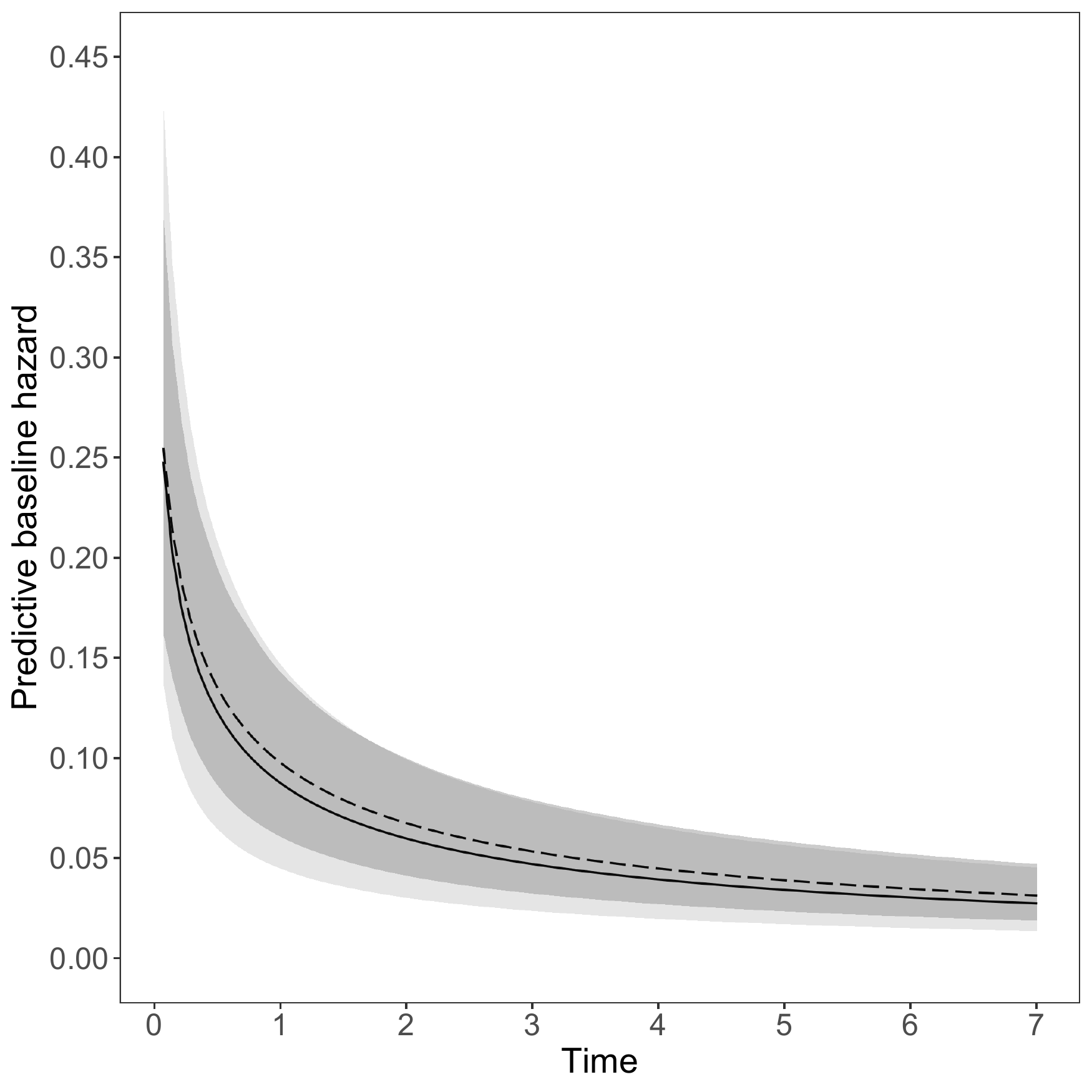}}}	
    \quad \quad \quad \quad
    \subfloat[]{{\includegraphics[width=7.5cm]{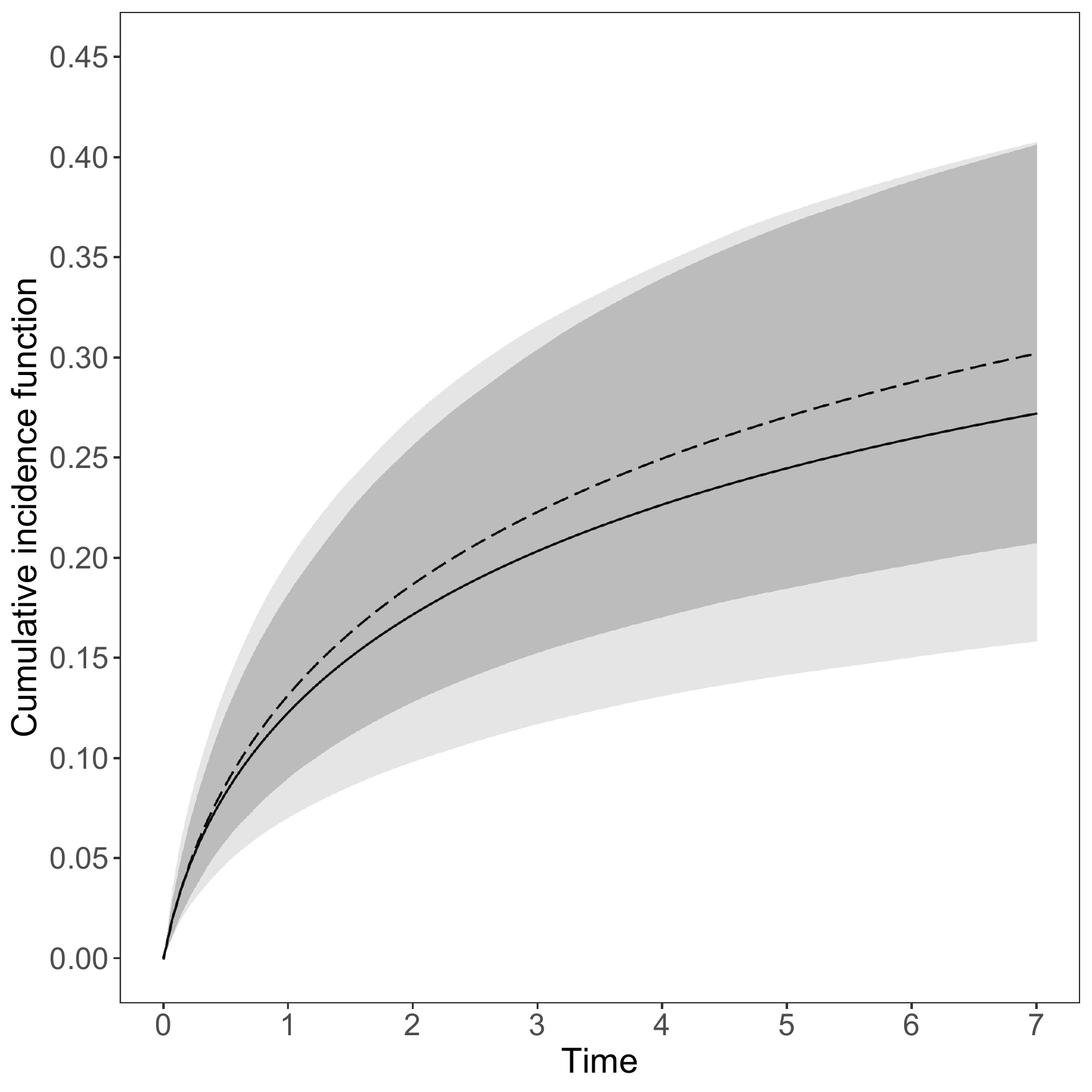}}}
    \caption{Log-normal predictive (a) baseline hazard and (b) cumulative incidence function of ISC (dashed line) and UAE (solid line), and their respective $95\%$ credible intervals (dark and light grey) for model $M_{2}$.}
    \label{fig:model2}
\end{figure}

These quantities can be approximated using Monte Carlo integration based on the posterior samples. Figure~\ref{fig:model2} shows the plots of \eqref{eq:predhaz} and \eqref{eq:cif} according to ISC and UAE risks, and their respective $95\%$ credible interval for model $M_{2}$.

The interpretations presented here are consistent with previous work that analysed this data \citep{williamson2007b,williamson2007a,williamson2008,hickey:2018}.

\section{Discussion} \label{sec:disc}

We have proposed a formulation of Bayesian joint models for longitudinal and survival data which allows for a relatively simple interpretation of the parameters and a tractable implementation. The idea is to model the survival process using a general hazard structure that separates the roles of the variables acting on the time scale from those that affect the hazard scale. This formulation can be coupled with the use of flexible parametric baseline hazards (e.g.~PGW or GG), which can capture a variety of hazard shapes, avoiding the need for numerical integration. 
We connect the survival process with the longitudinal process by sharing parameters with a similar interpretation. The longitudinal process can be modelled using GLMMs, allowing for the inclusion of a variety of response variables including continuous and categorical, within the Exponential family.
This formulation facilitates the implementation of the proposed joint models in a Bayesian framework using MCMC methods. In this paper, we have focused on the use of \texttt{rstan}, but other methods that allow for efficiently sampling from models with random effects can be used as well. 
We have presented a honest characterisation of the limitations of the proposed joint specification, which include guidelines on cases with high censoring rates, or with early administrative censoring. In such cases, the use of flexible parametric baseline hazards has to be taken with some care as, intuitively and as shown in our simulation study, there is not enough information to estimate the parameters controlling the tails. This is reflected on the resulting wide posterior distributions, compared to those associated to simpler choices of the baseline hazard (e.g.~log-normal). Model selection tools, such as Bayes factors or posterior model probabilities, are thus useful to identify the best model. In fact, the study of the performance of Bayesian model selection tools in the context of joint modelling represents a potential future research direction.
The real data applications presented here illustrate the flexibility of our formulation to be adapted to settings with competing or semi-competing risks in the survival process, adding another option to the toolbox for modelling these challenging scenarios \citep{andrinopoulou:2014,andrinopoulou:2017}. 
There are several natural extensions of the proposed joint models. For instance, other flexible parametric baseline hazards and flexible distributions on the longitudinal models can be employed instead of the ones presented here \citep{rubio:2018}.
Although we have presented a careful prior elicitation step, combining g-priors and weakly informative priors, we do not claim to have the last word on this point. Thus, other priors can be used as well (see  \cite{rossell:2019} for a discussion on different priors for survival models), and our R codes can be easily adapted for that purpose.

\section*{Acknowledgments}
FONDECYT (Chile), Grant/Award Number: 11190018.

\clearpage

\clearpage

\bibliographystyle{unsrtnat}
\bibliography{references}  

\end{document}